\documentclass[conference,a4paper]{IEEEtran}
\IEEEoverridecommandlockouts
\usepackage{cite}
\usepackage{amsmath,amssymb,amsfonts}
\usepackage{algorithm}
\usepackage{algpseudocode}
\usepackage{graphicx}
\usepackage{textcomp}
\usepackage{balance}
\usepackage{xcolor}
\usepackage{subfigure}
\usepackage{CJKutf8}
\def\BibTeX{{\rm B\kern-.05em{\sc i\kern-.025em b}\kern-.08em
    T\kern-.1667em\lower.7ex\hbox{E}\kern-.125emX}}

\newtheorem{theorem}{\textbf{Theorem}}

\begin{document}
\begin{CJK}{UTF8}{gbsn}

\title{Movable Antenna Array Aided Ultra Reliable Covert Communications}

\author{\IEEEauthorblockN{Yida Wang$^1$, Guojie Hu$^2$, Xiaoling Hu$^3$, Xingbo Lu$^4$, Yuzhen Huang$^4$}
\IEEEauthorblockA{$^1$College of Communications Engineering, Nanjing, 210007, China\\
$^2$College of Communication Engineering, Rocket Force University of Engineering, Xi${^{\prime}}$an 710025, China\\
$^3$State Key Laboratory of Networking and Switching Technology, \\Beijing University of Posts and Telecommunications, Beijing 100876, China\\
$^4$Academy of Military Science of PLA, Beijing 100080, China\\
Email: dadawang333@sina.com, lgdxhgj@sina.com, xiaolinghu@bupt.edu.cn, xingbo\underline{\hspace{0.5em}}lu@sina.com, yzh\underline{\hspace{0.5em}}huang@sina.com}

}

\maketitle

\begin{abstract}
In this paper, we construct a framework of the movable antenna (MA) aided covert communication shielded by the general noise uncertainty for the first time. According to the analysis performance on the derived closed-form expressions of the sum of the probabilities of the detection errors and the communication outage probability, the perfect covertness and the ultra reliability can be achieved by adjusting the antenna position in the MA array. Then, we formulate the communication covertness maximization problem with the constraints of the ultra reliability and the independent discrete movable position to optimize the transmitter's parameter. With the maximal
ratio transmitting (MRT) design for the beamforming, we solve the closed-form optimal information transmit power and design a lightweight discrete projected gradient descent (DPGD) algorithm to determine the optimal antenna position. The numerical results show that the optimal achievable covertness and the feasible region of the steering angle with the MA array is significant larger than the one with the fixed-position antenna (FPA) array.
\end{abstract}

\section{Introduction} \label{sec:1}
The secure communication can be guaranteed by the physical layer security (PLS) based on the property of the wireless medium to protect the communication content from eavesdropping  \cite{PLS_Wyner}. In order to improve the communication security, it is effective to exploit the spatial degree of freedom (DoF) achieved by the multiple antennas technology, such as the beamforming and the antenna selection \cite{PLS_multiple_antennas}. 
Compared with the conventional fixed-position antenna (FPA), the recently introduced movable antenna (MA) \cite{Lipeng_Magazine} can fully exploit the spatial DoF implemented by the local movement of antennas. Recently, it is proved that the communication security can be significantly improved with the help of the MA array \cite{Guojie_Secure}. Then, the secure communication with the MA array has been further investigated in the scenario without the eavesdropper (Eve)'s instantaneous channel state information (CSI) to combat with multiple non-collusive Eves \cite{Zhiyong_Feng} and collusive Eves \cite{Guojie_CSI}, respectively. In addition, the authors in \cite{Zan_Li_PLS} introduce the movable feature into the frequency diverse array to construct the movable frequency diverse antenna array (MFDA), which can be exploited to further improve the communication security.

It is noted that the aforementioned works focus on the protection of the communication content, rather than hiding the communication activity to meet high-level secure requirement. As a promising solution, covert communication aims to enable a communication while guaranteeing a negligible detection probability of this communication activity at a warden \cite{Bash}. Intuitively, the multiple antenna technology can be exploited to enhance the communication covertness. Specifically, the authors in \cite{covert_multiple_antennas_Tongxing} investigate the covert communication in a network with randomly located wardens and interferers, where the transmitter with a FPA array can adopt the maximal ratio transmitting (MRT) or distributed beamforming (DBF) strategy. The conclusion shows that the MRT strategy outperforms the DBF strategy in terms of the covert throughput. Moreover, the authors in \cite{covert_multiple_antennas_Jianquan} find that a positive covert rate is achievable in the delay-intolerant covert communication \cite{Delay-intolerant}, where the transmitter is equipped with a certain number of antennas in a FPA array and adopts the MRT strategy. In contrast to the FPA array, the MFDA has been also investigated in the delay-intolerant covert communication \cite{Zan_Li_Covert}, where the information transmission rate under the approximate communication covertness requirement can be significantly improved by the joint optimization of the beamforming, the frequency and the antenna position at the transmitter. Unfortunately, the essential impact of the MA on the exact communication covertness has to be further exploited.

Motivated by the effect of the MA aided interference cancellation, i.e., maximizing the received signal power over a desired direction and realizing the interference nulling over multiple undesired directions simultaneously \cite{Lipeng_null_steering}, we realize that the MA array has great potential to significantly improve the communication covertness. In addition, we note that the MA technology is suitable for the scenario with slowly varying channels \cite{Lipeng_Modelling}, especially for the machine-type communication (MTC) with ultra reliability requirement \cite{uRLLC}. Therefore, in this work, we investigate the MA array aided ultra reliable covert communication, where a transmitter equipped with a MA array intermittently transmits information to an intended receiver with the shield of the general noise uncertainty under the supervision of a warden with the detection ability for the communication activity. The main contributions of this work are summarized as follows.
\vspace{-1mm}
\begin{itemize}
 \item For the first time, we construct the framework of the MA aided covert communication shielded by the general noise uncertainty, i.e., the transmitter can adjust its beamforming, information transmit power and the antenna position to hide the communication activity from a warden’s detection under the noise uncertainty. Then, we analytically derive the sum of the probabilities of detection errors and the communication outage to evaluate the exact communication covertness and reliability, respectively. The performance analysis shows that either of the perfect covertness and the ultra reliability can be achieved by only adjusting the antenna position in the MA array.
\item We formulate an optimization problem to determine the beamforming factor, information transmit power $P_a$ and the antenna position $\mathbf{x}$ in the MA array to maximize the communication covertness subject to the ultra reliability requirement and the antenna movable limitation. After setting the beamforming factor as the rule of MRT to guarantee the ultra reliability, the formulated problem can be transformed to a new problem that $P_a$ and $\mathbf{x}$ are decoupled, and then they can be optimized separately. As such, we give the closed-form feasible optimal $P_a$, which only relates to the noise uncertainty, the number of the antenna, and the predetermined information transmission rate.
\item For the implement of the mechanical MA array in practice, we consider the independent movable region and the discrete optional positions for each antenna to avoid the potential mechanical collision and to match the characteristic of the stepping motor, respectively. Due to the highly non-convex property of the optimization of $\mathbf{x}$, we design a lightweight discrete projected gradient descent (DPGD) algorithm to determine the optimal $\mathbf{x}$ for the considered system. Compared with the conventional gradient descent algorithm, the DPGD algorithm needs to conduct 3 extra steps including the bounded, discretization, and the ultra reliability verification.
\end{itemize}

\textit{Notation}: Scalar variables are denoted by italic symbols.
Vectors are denoted by boldface symbols. Given a complex number, $\left|  \cdot  \right|$, ${\left(  \cdot  \right)^T}$, ${\left(  \cdot  \right)^\dag}$, and ${\left(  \cdot  \right)^H}$ denote the modulus, transpose, conjugate, and conjugate transpose (Hermitian), respectively. The minimize and maximize operations are represented as $\min \left( { \cdot \;, \cdot } \right)$ and $\max \left( { \cdot \;, \cdot } \right)$, respectively. Given an event, $\Pr \left(  \cdot  \right)$ denotes its probability of occurrence. $\mathbb{E}\left(  \cdot  \right)$ denotes the expectation operation. $\mathbb{C}^{a \times b}$ and $\mathbb{R}^{a \times b}$ denote the space of $a \times b$ complex-valued and real-valued matrices, respectively.

\section{System Model} \label{sec:2}

\begin{figure}
\begin{center}
  \includegraphics[width=3.2 in]{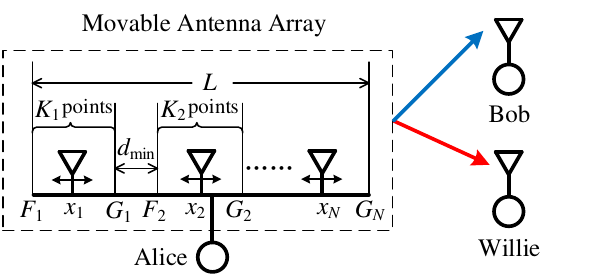}\\
\caption{The system model of covert communications aided by the movable antenna array.}
\end{center}
\vspace{-5mm}
\end{figure}

\subsection{Considered Scenario and Adopted Assumptions} \label{sec:2-1}
We consider a covert wireless communication system, where a transmitter (Alice) attempts to transmit confidential information covertly to a receiver (Bob) under the supervision of a warden (Willie), as shown in Fig. 1. Both of Bob and Willie are single-antenna devices, while Alice equipped with a linear MA array with $N$ antennas. The position of the $n$-th antenna at Alice is denoted as $x_n$, where $1 \le n \le N$. As such, the specific position of the MA array at Alice can be denoted as $\mathbf{x} = {\left[ {x_1,\;x_2, \cdots ,x_N} \right]^T} \in \mathbb{R}^{N \times 1}$. The movable range for the MA array at Alice is denoted as $[0, L={L_0}(N-1) {d_{\min }}]$, where $L_0$ is the movable scale factor, and $d_{\min}$ is the minimum distance between two adjacent antennas to avoid the coupling effect. Meanwhile, in order to avoid the potential MA mechanical collision, the movable range for each antenna at Alice is predetermined with a protection interval of $d_{\min}$. As such, the movable range can be expressed as ${x_n} \in [{F_n},{G_n}]$ \cite{Hu_Fluid}, where
\begin{align}
&{F_n}{\rm{ = }}\frac{{L - \left( {N - 1} \right){d_{\min }}}}{N}\left( {i - 1} \right){\rm{ + }}{d_{\min }}\left( {i - 1} \right), \label{eq0-1} \\
&{G_n}{\rm{ = }}\frac{{L - \left( {N - 1} \right){d_{\min }}}}{N}i{\rm{ + }}{d_{\min }}\left( {i - 1} \right). \label{eq0-2}
\end{align}
In practice, the optional movable position for each antenna at Alice is finite and discrete. Thus, we consider that the optional movable position for the $n$-th can be expressed as ${x_n} \in \left\{ {{x_n}\left( i \right)} \right\},\;i = 1,2, \cdots ,{K_n}$, where ${x_n(i)} \in [{F_n},{G_n}]$ and $K_n$ is the corresponding number of the optional movable position.

We assume that the communication activity is slotted. In addition, the beamforming factor at Alice is denoted as $\mathbf{w} \in \mathbb{C}^{N \times 1}$. Besides, the steering angles for Bob and Willie with respect to (w.r.t.) the MA array are denoted as $\theta_b$ and $\theta_w$, respectively. The channels of Alice-Bob and Alice-Willie are denoted as $\left\{ {\mathbf{h}_{ab}}{\rm{,}}{\mathbf{h}_{aw}} \right\} \in \mathbb{C}^{N \times 1} $, respectively. Considering the line-of-sight (LoS) propagation environment \cite{6DMA}, the corresponding channel vector is given by
\begin{align}\label{eq1}
{\mathbf{h}_{ai}} = {\left[ {{e^{j\frac{{2\pi }}{\lambda }{x_1}\cos {\theta _i}}},{e^{j\frac{{2\pi }}{\lambda }{x_2}\cos {\theta _i}}}, \cdots ,{e^{j\frac{{2\pi }}{\lambda }{x_N}\cos {\theta _i}}}} \right]^T},
\end{align}
where $i \in \{b,w\}$, $\lambda$ is the wavelength w.r.t. the information transmit frequency. When Alice transmits, the received signal at Bob is given by
\begin{align}\label{eq2}
{y_b} = \sqrt {{P_a}} \mathbf{h}_{ab}^H \mathbf{w}{x_a} + {n_b},
\end{align}
where $P_a$ is the information transmit power at Alice, $x_a \in \mathbb{C}$ is the information signal transmitted by Alice satisfying $\mathbb{E} \left( {{x_a}x_a^\dag } \right){\rm{ = }}1$, and $n_b\in \mathbb{C}$ is the additive white Gaussian noise (AWGN) at Bob with zero mean and variance $\sigma _b^2$. It is noted that the noise uncertainty exists at Bob, which can be characterized by the bounded noise uncertainty model proposed in \cite{HeBiao}. As such, the probability density function (PDF) of ${{\sigma }_{b}}$ is given by
\begin{align} \label{eq3}
{f_{\sigma _b^2}}\left( x \right) = \left\{ \begin{array}{l}
\frac{1}{{2\ln (\rho )x}},\;\;\;\frac{{\rm{1}}}{\rho }\sigma _0^2 \le x \le \rho \sigma _0^2{\rm{,}}\\
0{\rm{,}}\;\;\;\;\;\;\;\;\;\;\;\;\text{otherwise},
\end{array} \right.
\end{align}
where $\sigma _0^2$ is the nominal noise power, ${\sigma _b^2}$ is log-uniformly distributed range in the dB domain with the interval of $\left[ {\frac{1}{\rho }\sigma _0^2,\rho \sigma _0^2} \right]$, and $\rho$ is the parameter that quantifies the size of uncertainty \cite{Zhou}.

\subsection{Activity Detection at Willie} \label{sec:2-2}

In order to detect the potential communication activity between Alice and Bob from the perspective of Willie, Willie needs to conduct a binary hypothesis testing based on its observation within a time slot \cite{Bash}, in which Alice does not transmit in the null hypothesis $H_0$, but does transmit in the alternative hypothesis $H_1$.
Specifically, the composite received signal at Willie is given by
\begin{align}\label{eq4}
{y_w} = \left\{ {\begin{array}{*{20}{l}}
{{n_w},\;\;\;\;\;\;\;\;\;\;\;\;\;\;\;\;\;\;\;\;\;\;\;\;\;\;\;\;\;\;、；、；{H_0},}\\
{\sqrt {{P_a}} {\bf{h}}_{aw}^H{\bf{w}}{x_a} + {n_w},\;\;\;\;\;\;{H_1},}
\end{array}} \right.
\end{align}
where ${n}_w \in \mathbb{C}$ is the AWGN at Willie with zero mean and variance $\sigma _w^2$. Similarly, there also exists noise uncertainty at Willie, and the PDF of $\sigma_w$ is the same as \eqref{eq3}.

From a conservative point of view, we assume the channel vector of ${ {{\mathbf{h}_{ab}}} } $ and ${ {{\mathbf{h}_{aw}}} } $ are known by Willie. In addition, Willie can also acquire the PDFs of the noise, i.e.,  $f_{\sigma _b^2}\left( x \right)=f_{\sigma _w^2}\left( x \right)$, according to its statistical observation. Moreover, Willie has the ability to adopt the optimal strategy to determine whether there exists the communication activity between Alice and Bob by its received observations. Due to the existence of the noise uncertainty, Willie is hard to distinguish the variation of the noise with the existence of the communication activity, which will be further analyzed in the following section.

\section{Analysis on Performance Metrics for Ultra Reliable Covert Communications} \label{sec:3}

\subsection{Communication Covertness Performance} \label{sec:3-1}

The detection performance of Willie is normally measured
by the sum of the probabilities of detection errors \cite{Bash}, which is defined as
$\xi  = {P_{{\rm{FA}}}} + {P_{{\rm{MD}}}}$,
where ${P_\mathrm{FA}} = \Pr \left( {{D_1}{\rm{|}}{H_0}} \right)$ is the false alarm probability, ${P_\mathrm{MD}} = \Pr \left( {{D_0}{\rm{|}}{H_1}} \right)$ is the miss detection probability, $D_1$ and $D_0$ are the binary decisions indicating whether the Alice transmits or not, respectively.

According to the Neyman-Pearson criterion, the optimal strategy for the Willie to minimize its detection error probability is the likelihood ratio test (LRT), which can be transformed to the test of the average power received within a time slot \cite{LRT}. Specifically, such test is given by
\begin{align}\label{eq7}
{{T}_{w}}\underset{{{D}_{\text{0}}}}{\overset{{{D}_{\text{1}}}}{\mathop{\gtrless }}}\,\tau ,
\end{align}
where $T_w$ is the average power received at the Willie within a time slot, and $\tau$ is the Willie's detection threshold, which can be further optimized to minimize the sum of the probabilities of detection errors. Considering the case that one time slot contains sufficient channel use \cite{UAV_Relay}, as per \eqref{eq4}, $T_w$ is given by
\begin{align}\label{eq8}
{T_w} = \left\{ {\begin{array}{*{20}{l}}
{\sigma _w^2,\;\;\;\;\;\;\;\;\;\;\;\;\;\;\;\;\;\;\;\;\;\;\;\;{H_0},}\\
{{P_a}{{\left| {{\bf{h}}_{aw}^H{\bf{w}}} \right|}^2} + \sigma _w^2,\;\;\;{H_1}.}
\end{array}} \right.
\end{align}
\begin{theorem}
The Willie's minimum sum of the probabilities of detection errors is given by
\begin{align}\label{eq14}
{\xi ^ \ast } \!\!=\!\! \left\{ {\begin{array}{*{20}{l}}
{\!\!\!\!0,\;\;\;\;\;\;\;\;\;\;\;\;\;\;\;\;\;\;\;\;\;\;\;\;\;\;\;\;\;\;\;\;\;\;\;\;\;\;\;\;\;\;{P_a}{{\left| {{\bf{h}}_{aw}^H{\bf{w}}} \right|}^2} \ge {\kappa _0},}\\
{\!\!\!\!1 \!-\! \frac{{\rm{1}}}{{2\ln \left( \rho  \right)}}\ln \left( {1{\rm{ + }}\frac{{\rho {P_a}{{\left| {{\bf{h}}_{aw}^H{\bf{w}}} \right|}^2}}}{{\sigma _0^2}}} \right),{P_a}{{\left| {{\bf{h}}_{aw}^H{\bf{w}}} \right|}^2} \!\!<\!\! {\kappa _0},}
\end{array}} \right.
\end{align}
where $\kappa_0=\left( {\rho  - \frac{1}{\rho }} \right){\sigma _0^2}$.
\end{theorem}

\begin{IEEEproof}
As per \eqref{eq7} and \eqref{eq8}, when  ${P_a}{\left| {{\bf{h}}_{aw}^H{\bf{w}}} \right|^2} \ge \left( {\rho  - \frac{1}{\rho }} \right){\sigma _0^2}$, the expression of $\xi(\tau)$ is given by

\vspace{-2mm}

\begin{align}\label{eq12}
\xi \left( \tau  \right) = \left\{ {\begin{array}{*{20}{l}}
{1,\;\;\;\;\;\;\;\;\;\;\;\;\;\;\;\;\;\;\;\;\;\;\;\;\;\;\;\tau  \le \frac{{\sigma _0^2}}{\rho },}\\
{1 - \frac{{\ln \left( {\rho \tau /\sigma _0^2} \right)}}{{2\ln \left( \rho  \right)}},\;\;\;\;\;\;\;\;\frac{{\sigma _0^2}}{\rho } < \tau  < \frac{{\sigma _0^2}}{\rho } + {\kappa _1},}\\
{1 - \frac{{\ln \left( {\rho \tau /\sigma _0^2} \right)}}{{2\ln \left( \rho  \right)}} + \frac{{\ln \left( {(\tau  - {\kappa _{\rm{1}}})\rho /{\sigma _0^2}} \right)}}{{2\ln (\rho )}},\;}\\
{\;\;\;\;\;\;\;\;\;\;\;\;\;\;\;\;\;\;\;\;\;\;\;\;\;\;\;\;\;\frac{{\sigma _0^2}}{\rho } + {\kappa _1} \le \tau  \le \rho \sigma _0^2,}\\
{\frac{{\ln \left( {(\tau  - {\kappa _{\rm{1}}})\rho /{\sigma _0^2}} \right)}}{{2\ln (\rho )}},\;\;\;\;\;\rho \sigma _0^2 < \tau  < \rho \sigma _0^2 + {\kappa _1},}\\
{1,\;\;\;\;\;\;\;\;\;\;\;\;\;\;\;\;\;\;\;\;\;\;\;\;\;\;\tau  \ge \rho \sigma _0^2 + {\kappa _1}},
\end{array}} \right.
\end{align}
where $\kappa_1={P_a}{{{\left| {{\bf{h}}_{aw}^H{\bf{w}}} \right|}^2}}$.
Then, by the monotonicity analysis of $\xi(\tau)$ w.r.t. $\tau$ for \eqref{eq12} and the case with ${P_a}{\left| {{\bf{h}}_{aw}^H{\bf{w}}} \right|^2} < \left( {\rho  - \frac{1}{\rho }} \right){\sigma _0^2}$, we can complete the proof.
\end{IEEEproof}

As per \eqref{eq14}, when ${P_a}{\left| {{\bf{h}}_{aw}^H{\bf{w}}} \right|^2} \ge \left( {\rho  - \frac{1}{\rho }} \right){\sigma _0^2}$, the communication activity between Alice and Bob will be detected by Willie with no error. This illustrates that the smaller noise uncertainty is insufficient to shield the information transmission, i.e., the covert communications cannot be achieved. In contrast, when ${P_a}{\left| {{\bf{h}}_{aw}^H{\bf{w}}} \right|^2} < \left( {\rho  - \frac{1}{\rho }} \right){\sigma _0^2}$, $\xi^\ast$ can be improved with the decrease of $P_a$, i.e., the communication covertness with the MA array can still be improved by decreasing the information transmit power. In addition, when ${P_a}{\left| {{\bf{h}}_{aw}^H{\bf{w}}} \right|^2} < \left( {\rho  - \frac{1}{\rho }} \right){\sigma _0^2}$, we find that $\xi^\ast$ can be improved by setting proper channel gain of ${\left| {{\bf{h}}_{aw}^H{\bf{w}}} \right|^2}$. It is noted that the MA array has the ability to reconfigure the channel gain, hence there exists the potential to further enhance the communication covertness, which will be further exploited in the next section.

\subsection{Communication Outage Probability}\label{sec:3-2}

In this work, we consider a fixed-rate transmission from Alice to Bob, where $R$ is the predetermined transmission rate. Due to the existence of noise uncertainty at Bob, an event of communication outage from Alice to Bob occurs when $C \le R$, where $C={\log _2}\left( {1+{\gamma _b}} \right)$ is the channel achievable rate from Alice to Bob, ${\gamma _b} = \frac{{{P_a}{{\left| {{\bf{h}}_{ab}^H{\bf{w}}} \right|}^2}}}{{\sigma _b^2}}$ is the signal-to-noise ratio (SNR) at Bob. As such, the communication outage probability is denoted as $P_{\rm{out}}$, which is given by
\begin{align}\label{eq15}
 {{P}_{\rm{out}}}
 =\left\{ \begin{array}{l}
1{\rm{,}}\;\;\;\;\;\;\;\;\;\;\;\;\;\;\;\;\;\;\;\;{P_a}{\left| {{\bf{h}}_{ab}^H{\bf{w}}} \right|^2} \le \frac{{\sigma _0^2}}{\rho }\beta, \\
\ln \left( {\frac{{\rho \sigma _0^2\beta }}{{{P_a}{{\left| {{\bf{h}}_{ab}^H{\bf{w}}} \right|}^2}}}} \right){\rm{,}}\frac{{\sigma _0^2}}{\rho }\beta {\rm{ < }}{P_a}{\left| {{\bf{h}}_{ab}^H{\bf{w}}} \right|^2}{\rm{ < }}\rho \sigma _0^2\beta, \\
0{\rm{,}}\;\;\;\;\;\;\;\;\;\;\;\;\;\;\;\;\;\;\;\;{P_a}{\left| {{\bf{h}}_{ab}^H{\bf{w}}} \right|^2} \ge \rho \sigma _0^2\beta,
\end{array} \right.
\end{align}
where $\beta=2^R-1$.

As per \eqref{eq15}, when ${P_a}{\left| {{\bf{h}}_{ab}^H{\bf{w}}} \right|^2} \ge \rho \sigma _0^2\beta $, the communication outage probability is zero, i.e., the ultra reliability for the communication can be satisfied. It is obvious that increasing $P_a$ can achieve the ultra reliable communication more easily. Compared with the transmitter equipped with the FPA array, the transmitter equipped with the MA array can adjust the gain of ${\left| {{\bf{h}}_{ab}^H{\bf{w}}} \right|^2}$ to meet the ultra reliability condition more easily. In practice, the communication with high quality of service (QoS) often requires the ultra reliability, such as the scenario with the information transmitted from the implantable devices.

\section{Ultra Reliable Covert Communications Design} \label{sec:4}

As per \eqref{eq14} and \eqref{eq15}, the communication covertness and the ultra reliability condition are affected by the information transmit power and the setting of the MA array simultaneously. Considering the analysis in Section \ref{sec:3-1} and Section \ref{sec:3-2}, we then try to achieve the best trade-off between the communication covertness and reliability by optimizing the parameters at Alice, i.e., the beamforming factor $\mathbf{w}$, the information transmit power $P_a$, and the antenna position in the MA array $\mathbf{x} = {\left[ {x_1,\;x_2, \cdots ,x_N} \right]^T}$. Specifically, the optimization problem is given by
\begin{subequations}\label{eq16}
\begin{align}
({\rm{\mathbf{P1}}}):&{\rm{   }}\mathop \mathrm{maximize}\limits_{\mathbf{w},\;P_a,\;\mathbf{x}} \;\; \xi^\ast  \\
&{\rm{            }}\;{\rm{s.t.}}\quad {P_a}{\left| {{\bf{h}}_{ab}^H{\bf{w}}} \right|^2} \ge \rho \sigma _0^2\beta , \label{eq16a}\\
&{\rm{            }}\quad \quad\;\; {x_n} \in \left\{ {{x_n}\left( i \right)} \right\},\label{eq16b}\\
&{\rm{            }}\quad \quad\;\; {x_n(i)} \in [{F_n},{G_n}],\label{eq16c}\\
&{\rm{            }}\quad \quad\;\; n=1,2, \cdots ,{N},\label{eq16d}\\
&{\rm{            }}\quad \quad\;\; i = 1,2, \cdots ,{K_n}, \label{eq16e}
\end{align}
\end{subequations}
where \eqref{eq16a} is the constraint of the ultra reliability requirement, \eqref{eq16b}-\eqref{eq16e} are the movable limitations for the MA array. In addition, we recall that $\beta=2^R-1$, $F_n$ and $G_n$ can refer to \eqref{eq0-1} and \eqref{eq0-2}, respectively.

In order to guarantee the ultra reliable communication between Alice and Bob, we consider the maximum ratio transmission (MRT) at Alice \cite{MRT}. As such, the beamforming factor at Alice is given by $\mathbf{w}{\rm{ = }}\frac{{{\mathbf{h}_{ab}}}}{{\left| {{\mathbf{h}_{ab}}} \right|}}$. As per \eqref{eq1}, we note that ${\left| {{\mathbf{h}_{ab}}} \right|^2}{\rm{ = }}{N}$. Thus, as per the expression of $\xi^\ast$ in \eqref{eq14}, the optimization problem \eqref{eq16} can be transformed as follows.
\begin{subequations}\label{eq18}
\begin{align}
({\rm{\mathbf{P2}}}):&{\rm{   }}\mathop \mathrm{minimize}\limits_{P_a,\;\mathbf{x}} \;\; {{P_a}{{\left| {{\bf{h}}_{aw}^H{\mathbf{h}_{ab}}} \right|}^{\rm{2}}}} \label{eq18a} \\
&{\rm{            }}\;{\rm{s.t.}}\quad \frac{{{P_a}{{\left| {{\mathbf{h}}_{aw}^H{\mathbf{h}_{ab}}} \right|}^{\rm{2}}}}}{{{N}}} \le \left( {\rho  - \frac{1}{\rho }} \right){\sigma _0^2}, \label{eq18b}\\
&{\rm{            }}\quad \quad\;\; {P_a} N \ge \rho \sigma _0^2\beta, \label{eq18c} \\
&{\rm{            }}\quad \quad\;\; \mathbf{x} \in {\cal C},\label{eq18d}
\end{align}
\end{subequations}
where $\cal C$ is the feasible region for $\mathbf{x}$ with the constraints \eqref{eq16b}-\eqref{eq16e}.

According to the optimization problem \eqref{eq18}, the corresponding optimization object is decoupled w.r.t. $P_a$ and $\mathbf{x}$. Thus, the optimization problem \eqref{eq18} can be solved w.r.t. $P_a$ and $\mathbf{x}$ sequentially. For a given $\mathbf{x}$, it is obvious that the optimization of $P_a$ is convex and the corresponding solution is given by
\begin{align}\label{eq20}
P_a^{\rm{\dag}}{\rm{ = }}\left\{ \begin{array}{l}
\frac{{\beta \rho \sigma _0^2}}{{{N}}}{\rm{,}}\; {\left| {{\mathbf{h}}_{aw}^H{{\mathbf{h}}_{ab}}} \right|^{\rm{2}}} \le \left( {1 - \frac{1}{{{\rho ^2}}}} \right)\frac{{{N^2}}}{\beta }{\rm{,}}\\
\;\phi,\;\;\;\;\;\;\text{otherwise},
\end{array} \right.
\end{align}
where $\phi$ illustrates the corresponding optimization problem is unsolvable. By substituting \eqref{eq20} into \eqref{eq18a}-\eqref{eq18c}, the optimization problem \eqref{eq18} w.r.t. $\mathbf{x}$ can be formulated as follows.
\vspace{-5mm}
\begin{subequations}\label{eq21}
\begin{align}
({\rm{\mathbf{P3}}}):&{\rm{   }}\mathop \mathrm{minimize}\limits_{\mathbf{x}} \;\; f_0(\mathbf{x})  \\
&{\rm{            }}\;{\rm{s.t.}}\quad f_0(\mathbf{x}) \le \left( {1 - \frac{1}{{{\rho ^2}}}} \right)\frac{{{N^2}}}{\beta }, \label{eq21a}\\
&{\rm{            }}\quad \quad\;\; \mathbf{x} \in {\cal C},\label{eq21b}
\end{align}
\end{subequations}
where ${f_0}({\mathbf{x}}) = {\left| {{\mathbf{h}}_{aw}^H{{\mathbf{h}}_{ab}}} \right|^{\rm{2}}}$. We note that the optimization problem \eqref{eq21} is highly non-convex due to the complex objective. To tackle this problem, the DPGD algorithm, i.e., the discrete projected gradient descent algorithm, can be exploited to find the local optimal solution. Specifically, the update rule for $\mathbf{x}$ is given by
\begin{align}\label{eq22}
{\mathbf{x}^{t{\rm{ + }}1}} &= {\cal D}\left( \mathbf{\tilde x}^{t+1},{\cal C} \right) = {\cal D}\left( {{\cal B}\left( {{\mathbf{\bar {x}}^{t+1 }},{\cal C}} \right)},{\cal C} \right) \nonumber \\
& = {\cal D}\left( {{\cal B}\left( {{{\mathbf{x}}^t} - \delta {\nabla _t}{f_0}\left( {{{\mathbf{x}}^t}} \right),{\cal C}} \right),{\cal C}} \right),
\end{align}
where ${{\mathbf{x}}^t}= \left( {  x_1^{t }, x_2^{t }, \cdots , x_N^{t }} \right)$ is the updated $\mathbf{x}$ in the $t$-th iteration, $ \mathbf{{\bar x}}^{t+1} = \left( { \bar x_1^{t + 1},\bar x_2^{t + 1}, \cdots ,\bar x_N^{t + 1}} \right)$ and
$ \mathbf{\tilde{x}}^{t+1} = \left( {\tilde{x}_1^{t + 1},\tilde{x}_2^{t + 1}, \cdots ,\tilde{x}_N^{t + 1}} \right)$ are two auxiliary variables in the $(t+1)$-th iteration, $\delta $ is the step size for gradient descent, and ${\nabla _t}{f_0}\left( {{{\bf{x}}^t}} \right)$ is the gradient of $f_0(\mathbf{x})$ w.r.t. $\mathbf{x}^t$. In addition, ${\cal B}\left( \cdot,\cdot \right)$ is the bounded operation for the gradient descent, and ${\cal D}\left( \cdot,\cdot \right)$ is the discretized operation. As shown in Algorithm 1., the specific step for the DPGD method for the optimization problem \eqref{eq21} can be described as follows.

\begin{figure*}[t]
\centering
\subfigure[]{
\label{Fig.sub.1-1}
\includegraphics[width=0.32\textwidth]{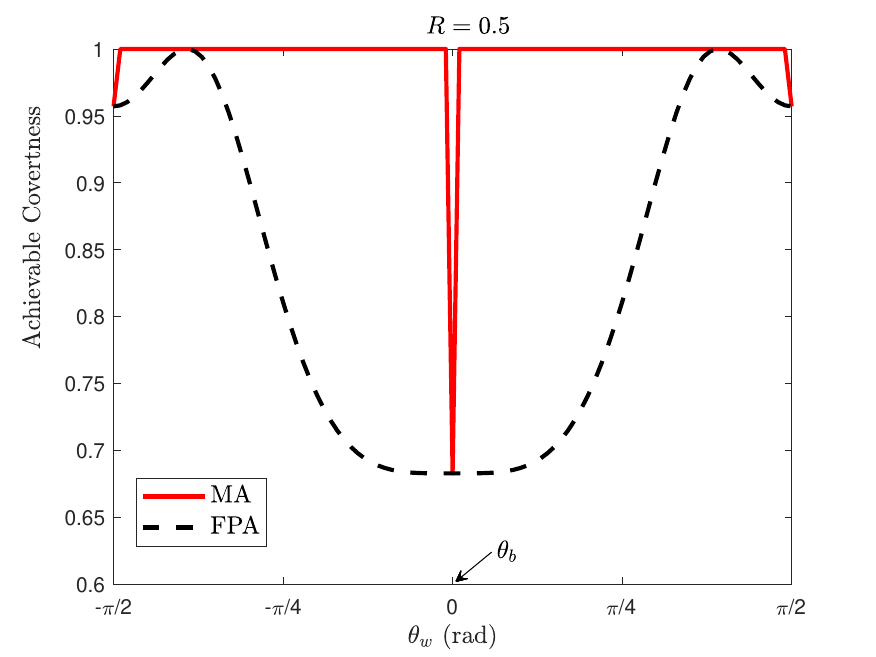}}
\subfigure[]{
\label{Fig.sub.1-2}
\includegraphics[width=0.32\textwidth]{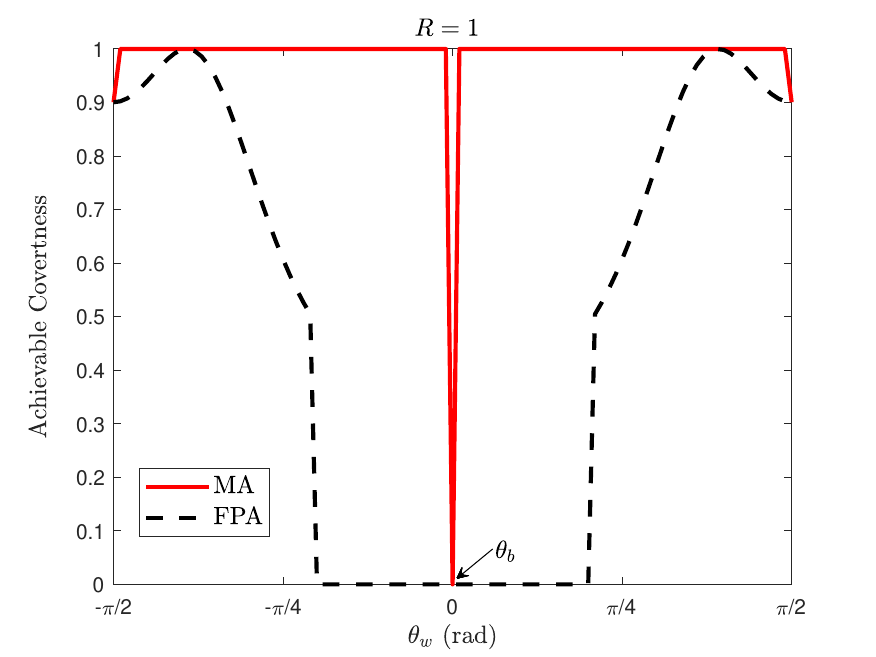}}
\subfigure[]{
\label{Fig.sub.1-3}
\includegraphics[width=0.32\textwidth]{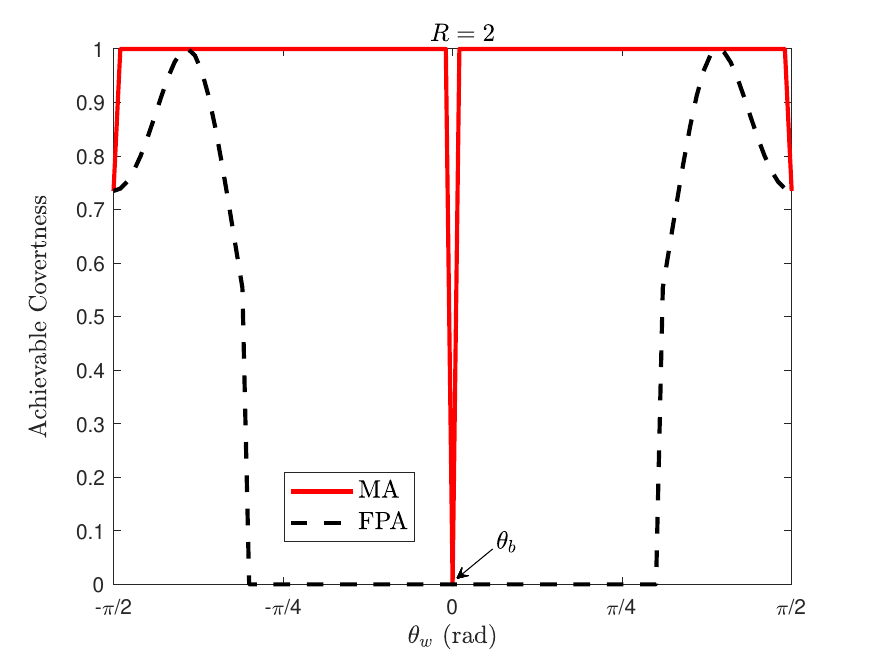}}
\caption{The achievable covertness versus the Willie's steering angle $\theta_w$  for different $R$ with $N\!=\!3$, $K_n\!=\!+\infty $, $L_0=10$ and $\rho=3$ dB.}
\label{Fig2}
\vspace{-5mm}
\end{figure*}

$\emph{1)}$ \textit{Conduct the gradient descent}:

First, we derive ${\nabla _t}{f_0}\left( \mathbf{x}^t \right)$ as follows.
\begin{align}\label{eq23}
{\nabla _t}{f_0}\left( \mathbf{x}^t \right) & = \left( {\frac{{\partial \left( {{f_0}\left( \mathbf{x}^t \right)} \right)}}{{\partial {x_1^t}}},\frac{{\partial \left( {{f_0}\left( \mathbf{x}^t \right)} \right)}}{{\partial {x_2^t}}}, \cdots ,\frac{{\partial \left( {{f_0}\left( \mathbf{x}^t \right)} \right)}}{{\partial {x_N^t}}}} \right) \nonumber \\
& = \left( {\varphi \left( {{x_1^t}} \right){\rm{,}}\varphi \left( {{x_2^t}} \right){\rm{,}} \cdots ,\varphi \left( {{x_N^t}} \right)} \right),
\end{align}
where $\varphi \left( {{x_n^t}} \right)\!=\!2A\left( {{T_1}\cos \left( {A{x_i^t}} \right) \!-\! {T_2}\sin \left( {A{x_n^t}} \right)} \right)$, ${T_1}\!\!=\!\!\!\!\sum\limits_{n{\rm{ = }}1}^N {\sin \left( {A{x_n^t}} \right)} $, ${T_2}\!\!=\!\!\!\!\sum\limits_{n{\rm{ = }}1}^N {\cos \left( {A{x_n^t}} \right)} $, $A\!\!=\!\!\frac{{2\pi }}{\lambda }\left( {\cos {\theta _b}\!+\!\cos {\theta _w}} \right)$. Then, $\mathbf{{\bar x}}^{t+1}$ is given by
\begin{align}\label{eq24}
\mathbf{{\bar x}}^{t+1}=  {{\mathbf{x}}^t} - \delta {\nabla _t}{f_0}\left( {{{\mathbf{x}}^t}} \right).
\end{align}

$\emph{2)}$ \textit{Conduct the bounded operation, i.e.,} ${\cal B}\left( {{{{\bf{\bar x}}}^{t + 1}},{\cal C}} \right)$:
\begin{align}\label{eq25}
{{\tilde x}_n}^{t + 1} = \max \left( {\min \left( {{{\bar x}_n}^{t + 1},{G_n}} \right),{F_n}} \right),
\end{align}
where we recall that $F_n$ and $G_n$ are the lower position bound and upper position bound for $x_n$, respectively, and the corresponding values are given as \eqref{eq0-1} and \eqref{eq0-2}.

$\emph{3)}$ \textit{Conduct the discretization, i.e.,} ${\cal D}\left( {{{{\bf{\tilde x}}}^{t + 1}},{\cal C}} \right)$:
\begin{align}\label{eq26}
x_n^{t+1}=x_n(k^{t+1}_n),
\end{align}
where $k_n^{t+1} = \mathop {\arg \min }\limits_{{k_n} = 1,2, \cdots ,{K_n}} \left| {x_n\left( {{k_n}} \right) - \tilde x_n^{t + 1}} \right|$.

$\emph{4)}$ \textit{Ultra reliability verification}:

The DPGA method iteratively updates $\mathbf{x}^{t+1}$ based on \eqref{eq22} until $\left| {{f_0}\left( {{{\mathbf{x}}^{t{\rm{ + }}1}}} \right) - {f_0}\left( {{{\mathbf{x}}^t}} \right)} \right| \le \epsilon$, where $\epsilon $ is the a tiny tolerance constant value. When $f_0(\mathbf{x}^{t+1}) \le \left( {1 - \frac{1}{{{\rho ^2}}}} \right)\frac{{{N^2}}}{\beta }$, the optimal antenna position for the MA array is given by
\begin{align}\label{eq27}
\mathbf{x^\ast}=\mathbf{x}^{t+1}=\left( {  x_1^{t+1 }, x_2^{t+1 }, \cdots , x_N^{t+1 }} \right).
\end{align}
It is noted that the considered system cannot achieve the ultra reliability when $f_0(\mathbf{x}^{t+1}) > \left( {1 - \frac{1}{{{\rho ^2}}}} \right)\frac{{{N^2}}}{\beta }$.

$\emph{5)}$ \textit{Conduct the ultra reliable covert communication aided by the MA array:}

After the ultra reliability verification, the optimal information transmit power can be given by
\begin{align}\label{eq28}
P_a^{\ast} = \frac{{\beta \rho \sigma _0^2}}{{{N}}}.
\end{align}
And the corresponding beamforming factor $\mathbf{w}^{\ast}$ can be acquired by substituting $\mathbf{x}^{\ast}$ into $\mathbf{w}=\frac{{{\mathbf{h}_{ab}}}}{{\left| {{\mathbf{h}_{ab}}} \right|}}$. In practice, each antenna in MA array can be mechanically moved simultaneously by the instructions based on $\mathbf{x}^{\ast}$. Meanwhile, the information transmit power and the beamforming factor at Alice can set as $P_a^\ast$ and $\mathbf{w}^\ast$, respectively.

  \begin{algorithm}[htb]
  \caption{The DPGD Algorithm for the Ultra Reliable Covert Communication Aided by the MA Array}
  \begin{algorithmic}[1]
 \State \textbf{Input:}
$\theta_b$, $\theta_w$, $N$, $K_n$, $L_0$, $d_{\min}$, $\rho$, $\sigma_0^2$, $R$, $\delta$, $\epsilon$.
 \State Initialization: ${\mathbf{x}^0}{\rm{ = }}\left( {{F_1}{\rm{,}}{F_2}{\rm{,}} \cdots {\rm{,}}{F_N}} \right)$
 \State \textbf{while} $\left| {{f_0}\left( {{{\bf{x}}^{t{\rm{ + }}1}}} \right) - {f_0}\left( {{{\bf{x}}^t}} \right)} \right| \le \epsilon$ \textbf{do}
 \State \;\;\;Update $\mathbf{{\bar x}}^{t+1}$ based on \eqref{eq24}.
 \State \;\; Conduct ${\cal B}\left( {{{{\bf{\bar x}}}^{t + 1}},{\cal C}} \right)$ based on \eqref{eq25}.
  \State \;\; Conduct ${\cal D}\left( {{{{\bf{\tilde x}}}^{t + 1}},{\cal C}} \right)$ based on \eqref{eq26}.
  \State \textbf{end while}
  \State \textbf{if}\;$f_0(\mathbf{x}^{t+1}) \le \left( {1 - \frac{1}{{{\rho ^2}}}} \right)\frac{{{N^2}}}{\beta }$ \textbf{do}
  \State \;\;\;Acquire $\mathbf{x}^{\ast}$ based on \eqref{eq27}.
  \State \;\;\;Acquire $P_a^{\ast}$ based on \eqref{eq28}.
  \State  \;\;\;Acquire $\mathbf{w}^{\ast}$ by substituting $\mathbf{x}^{\ast}$ into $\mathbf{w}=\frac{{{\mathbf{h}_{ab}}}}{{\left| {{\mathbf{h}_{ab}}} \right|}}$.
  \State \textbf{end if }\;
  \State Return $\mathbf{x}^{\ast}$, $P_a^{\ast}$, $\mathbf{w}^{\ast}$.
  \end{algorithmic}
  \end{algorithm}

\section{Numerical Results} \label{sec5}
In this section, numerical results are presented to demonstrate the effectiveness of the MA array for the ultra reliable covert communications. We set the minimum distance between the adjacent antennas in the MA array as $d_{\min}=\lambda/2$, the noise power as $\sigma_0^2 = 1$ for normalizing the large-scale channel fading power, the steering angle at Bob as $\theta_b=0$, the step size for the DPGD method as $\delta = 0.001$, and the tolerance constant value as $\epsilon=10^{-5}$. As a benchmark, we also examine the performance in the case with the FPA array. For performance comparison, we optimize the information transmit power $P_a$ and set the beamforming factor as MRT in the case with the FPA array.

In Fig. 2, we plot the achievable covertness, i.e., the optimal $\xi^\ast$ in the optimization problem \eqref{eq16}, versus the steering angle at Willie $\theta_w$ (rad) for different $R$. In this figure, we first observe that the achievable covertness with the MA array approximates to 1 stably, i.e., almost perfect covertness, except when $\theta_w$ approaches to the value of $\theta_b$ or $\theta_b \pm \pi/2$, while the achievable covertness with the FPA array undergoes rapid variation w.r.t. $\theta_w$. Then, we observe that with larger $R$, the transmitter equipped with the MA array can still achieve the almost perfect covertness for most $\theta_w$, while the covert communication is infeasible for the transmitter equipped with the FPA array for some $\theta_w$ approaching to $\theta_b$. The aforementioned two observations demonstrate the benefits of the application of the MA array in the ultra reliable covert communication.

\begin{figure}
\centering
\label{Fig.3}
\includegraphics[width=0.4\textwidth]{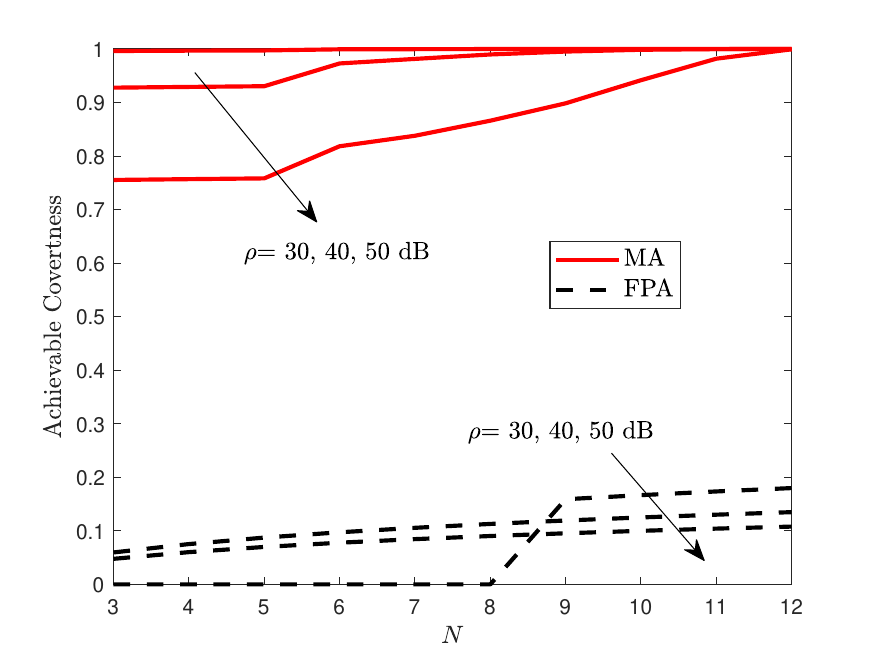}
\vspace{-4mm}
\caption{The achievable covertness versus the number of the antenna $N$ for different $\rho$ with $\theta_w=0.01\pi/2$, $K_n=+\infty $, $L_0=2$, and $R=1$.}
\vspace{-3mm}
\end{figure}

\begin{figure}
\centering
\label{Fig.4}
\includegraphics[width=0.4\textwidth]{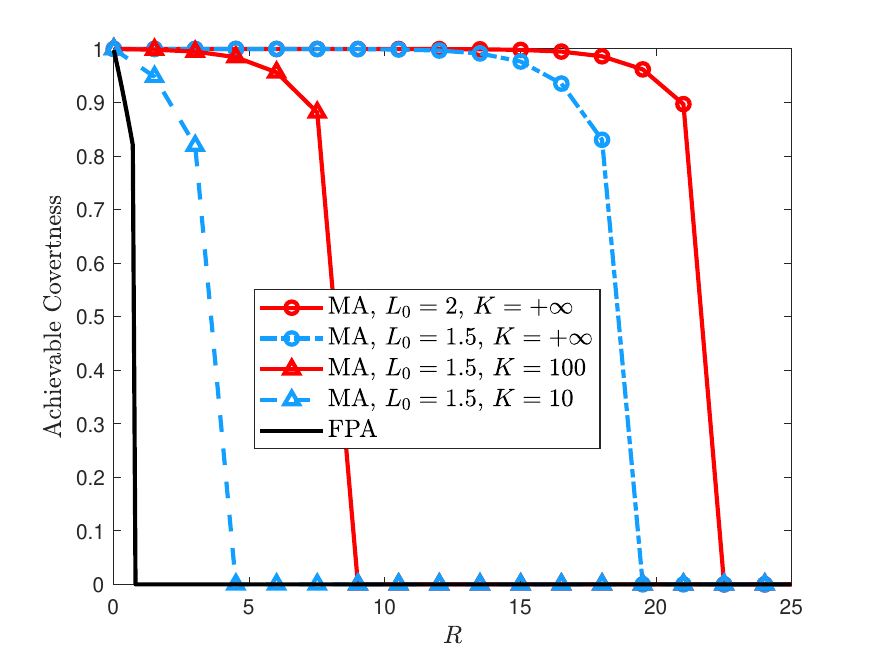}
\vspace{-4mm}
\caption{The achievable covertness versus the transmission rate $R$ for different $L_0$ and $K_n=K$ with $\theta_w=0.01\pi/2$, $N=4$, and $\rho=3$ dB.}
\vspace{-5mm}
\end{figure}

In Fig. 3, we plot the achievable covertness versus the number of the antenna $N$ for different $\rho$. In this figure, we first observe that the achievable covertness with the MA array is significantly higher than the one with the FPA array, even when the ultra reliable communication is infeasible in the case with $\rho=50$ dB and $N<8$ for configuration with the FPA array. Then, we observe that both of the achievable covertness with the MA array and the one with the FPA array decrease with the increase of $\rho$, i.e., the noise uncertainty size, which is caused by the negative effect on the ultra reliability.

In Fig. 4, we plot the achievable covertness versus the transmission rate $R$ for different $L_0$ and $K_n=K$. In this figure, we first observe that the both of the achievable covertness with the MA array and the one with the FPA array decrease as the increase of $R$. Moreover, we observe that the achievable covertness with the FPA array can acquire a larger value w.r.t. larger $L_0$ and $K$. This illustrates that the larger movable region and the discretization with more levels can acquire larger spatial gain for the covertness with the MA.

\section{Conclusion} \label{sec5}
In this paper, we construct a framework of the MA aided covert communication shielded by the general noise uncertainty. To evaluate the communication covertness and reliability, we analytically derive the sum of the probabilities of the detection errors and the communication outage probability, respectively. The performance analysis reveals that either of the perfect covertness and the ultra reliability can be achieved by the adjustment of the antenna position in the MA array. Then, we formulate the covertness maximization problem under the constraints of the ultra reliability and the movable position to determine the transmitter's parameters. We give the specific setting guidance for the beamforming factor and the information transmit power, as well as a lightweight DPGD algorithm to adjust the antenna position in the MA array. The numerical results show that the optimal achievable covertness and the feasible region of the steering angle with the MA array are significant larger than the one with the FPA array.

\section{Acknowledgment} \label{sec6}
This work is supported by the National Natural Science Foundation of China under Grant 62201606.

\balance

\end{CJK}

\begin{thebibliography}{00}
\vspace{-1mm}
\bibitem{PLS_Wyner}
A. D. Wyner, ``The wire-tap channel,'' \emph{The Bell System Technical Journal}, vol. 54, no. 8, pp. 1355-1387, Oct. 1975.

\bibitem{PLS_multiple_antennas}
X. Chen, D. W. K. Ng, W. H. Gerstacker, and H. H. Chen, ``A survey on multiple-antenna techniques for physical layer security,'' \emph{IEEE Commun Surv. Tut.}, vol. 19, no. 2, pp. 1027-1053, Secondquarter 2017.

\bibitem{Lipeng_Magazine}
L. Zhu, W. Ma, and R. Zhang, ``Movable antennas for wireless communication: Opportunities and challenges,'' \emph{IEEE Commun. Mag.}, vol. 62, no. 6, pp. 114-120, Jun. 2024.


\bibitem{Guojie_Secure}
G. Hu, Q. Wu, K. Xu, J. Si, and N. Al-Dhahir, ``Secure wireless communication via movable-antenna array,'' \emph{IEEE Signal Process. Lett.}, vol. 31, pp. 516-520, Jan. 2024.

\bibitem{Zhiyong_Feng}
Z. Feng, Y. Zhao, K. Yu, and D. Li. ``Movable antenna empowered physical layer security without eve's CSI: Joint optimization of beamforming and antenna positions,'' \emph{arXiv preprint}, arXiv:2405.16062, May 2024.

\bibitem{Guojie_CSI}
G. Hu, Q. Wu, D. Xu, K. Xu, J. Si, Y. Cai, et al., ``Movable antennas-assisted secure transmission without eavesdroppers' instantaneous CSI,'' \emph{IEEE Trans Mob. Comput.}, Early Access, DOI: 10.1109/TMC.2024.3438795, Aug. 2024.



\bibitem{Zan_Li_PLS}
Z. Cheng, J. Si, Z. Li, P. Liu, Y. Huang, and N. Al-Dhahir, ``Movable frequency diverse array for wireless communication security,'' \emph{arXiv preprint}, arXiv:2407.21157, Jul. 2024.


\bibitem{Bash}
B. A. Bash, D. Goeckel, and D. Towsley, ``Limits of reliable communication with low probability of detection on AWGN channels," \emph{IEEE J. Sel. Areas Commun.}, vol. 31, pp. 1921-1930, Sep. 2013.



\bibitem{covert_multiple_antennas_Tongxing}
T. Zheng, H. Wang, D. W. K. Ng, and J. Yuan, ``Multi-antenna covert communications in random wireless networks,'' \emph{IEEE Trans. Wireless Commun.}, vol. 18, no. 3, pp. 1974-1987, Mar. 2019.



\bibitem{covert_multiple_antennas_Jianquan}
W. Xiang, J. Wang, S. Xiao, and W. Tang, ``Achieving constant rate covert communication via multiple antennas,'' in \emph{Proc. VTC2022-Spring}, Jun. 2022, pp. 1-6.

\bibitem{Delay-intolerant}
S. Yan, B. He, X. Zhou, Y. Cong, and A. L. Swindlehurst, ``Delay-intolerant covert communications with either fixed or random transmit power,'' \emph{IEEE Trans.Inf. Forensics Security}, vol. 14, no. 1, pp. 129-140, Jan. 2019.


\bibitem{Zan_Li_Covert}
Z. Cheng, J. Si, Z. Li, P. Liu, X. Wang, and N. Al-Dhahir, ``Movable frequency diverse array-assisted covert communication with multiple wardens,'' \emph{arXiv preprint}, 	arXiv:2407.20280, Jul. 2024.


\bibitem{Lipeng_null_steering}
L. Zhu, W. Ma, and R. Zhan, ``Movable-antenna array enhanced beamforming: Achieving full array gain with null steering,'' \emph{IEEE Commun. Lett.}, vol. 27, no. 12, pp. 3340–3344, Oct. 2023.


\bibitem{Lipeng_Modelling}
L. Zhu, W. Ma, and R. Zhang, ``Modeling and performance analysis for movable antenna enabled wireless communications,'' \emph{IEEE Trans. Wireless Commun.}, vol. 23, no. 6, pp. 6234-6250, Jun. 2024.



\bibitem{uRLLC}
C. She, C. Pan, T. Q. Duong, T. Q. S. Quek, R. Schober, M. Simsek, et al., ``Guest editorial xURLLC in 6G: Next generation ultra-reliable and low-latency communications,'' \emph{IEEE J. Select. Areas Commun.}, vol. 41, no. 7, pp. 1963-1968, Jul. 2023.


\bibitem{Hu_Fluid}
G. Hu, Q. Wu, K. Xu, J. Ouyang, J. Si, Y. Cai, et al., ``Fluid antennas-enabled multiuser uplink: A low-complexity gradient descent for total transmit power minimization,'' \emph{IEEE Commun. Lett.}, vol. 28, no. 3, pp. 602-606, Mar. 2024.

\bibitem{6DMA}
X. Shao, Q. Jiang, and R. Zhang. ``6D movable antenna based on user distribution: Modeling and optimization,'' \emph{arXiv preprint}, arXiv:2403.08123, Mar. 2024.


\bibitem{HeBiao}
B. He, S. Yan, X. Zhou, and V. K. N. Lau, ``On covert communication with noise uncertainty,'' \emph{IEEE Commun. Lett.}, vol. 21, no. 4, pp. 941-944, Apr. 2017.


\bibitem{Zhou}
X. Zhou, S. Yan, J. Hu, J. Sun, J. Li, and F. Shu, ``Joint optimization of a UAV's trajectory and transmit power for covert communications,'' \emph{IEEE Trans. Signal Process.}, vol. 67, no. 16, pp. 4276-4290, Aug. 2019.

\bibitem{LRT}
R.~Chen, Z.~Li, J.~Shi, L.~Yang, and J.~Hu, ``Achieving covert communication in overlay cognitive radio networks,'' \emph{IEEE Trans. Veh. Technol.}, vol.~69, no.~12, pp. 15\,113--15\,126, Dec. 2020.

\bibitem{UAV_Relay}
M. Li, X. Tao, H. Wu and N. Li, ``Joint trajectory and resource optimization for covert communication in UAV-enabled relaying systems,'' \emph{IEEE Trans. Veh. Technol.}, vol. 72, no. 4, pp. 5518-5523, Apr.2023.

\bibitem{MRT}
T. K. Y. Lo, ``Maximum ratio transmission,'' \emph{IEEE Trans. Commun.}, vol. 47, no. 10, pp. 1458-1461, Oct. 1999.




\end{thebibliography}
\end{document}